%% file: slacconf_one.tex
\documentclass[slac_one]{revtex4}
\usepackage{graphicx}
\usepackage{fancyhdr}
\input ./colordvi.fil 
\input ./symbols_util.fil 
\input ./symbols_physics.fil 
\input ./symbols_particles.fil

\begin{document}

\title{{\small{Hadron Collider Physics Symposium (HCP2008),
Galena, Illinois, USA}}\\ %% Please keep this conference title here
\vspace{12pt}
Search for the $Wh$ Production Using High-$p_{T}$ Isolated Like-Sign Dilepton 
Events in Run-II with 1.9 ${\rm fb^{-1}}$} %% Paper title goes here


\author{Toru Okusawa, Yoshihiro Seiya, Takayuki Wakisaka and 
        Kazuhiro Yamamoto}
\affiliation{Osaka City University, Japan, on behalf of the CDF Collaboration}

\begin{abstract}
We search for the neutral higgs production associated with the $W$ boson using
high-$p_{T}$ isolated like-sign dilepton events in $p\bar{p}$ collisions at 
$\sqrt{s}$ = 1.96 TeV. The data were collected with the CDF-II detector at the 
Fermilab Tevatron collider and correspond to an integrated luminosity of 1.9
${\rm fb}^{-1}$. We examine the dilepton events on the plane of the 2nd lepton 
$p_{T}$ ($p_{T2}$) versus the $ p_{T}$ of dilepton system ($p_{T12}$)
and tune the final cut to maximize the sensitivity on the plane.
We set the final cut to the 2nd lepton $p_{T}$ greater than 20 GeV/c 
and the $p_{T}$ of dilepton system greater than 15 GeV/c. 
The expected number of signal events is 0.46 for the fermiophobic higgs 
of the mass 110 GeV/$c^{2}$ 
and 0.19 for the mass 160 GeV/$c^{2}$ 
assuming the Standard Model cross section.
The number of events for the Standard Model higgs is 0.02 and 0.18 for 
110 GeV/$c^2$ and 160 GeV/$c^{2}$, respectively. The expected number of
backgrounds is 3.23 $\pm$ 0.69, while we observed 3 events in the data. 
From these results, we obtain the limits on $\sigma(p\bar{p}\rightarrow Wh)
\times Br(h\rightarrow W^{+}W^{-})$ 
of 2.2 pb for the higgs of 
110 GeV/$c^{2}$ and 1.4 pb for 160 GeV/$c^2$ at the 95$\%$ confidence level.
\end{abstract}

\maketitle

\thispagestyle{fancy}


%------------------------------------------------------------------------------
% Introduction 
%------------------------------------------------------------------------------
\section{\label{sec:Intro}Introduction}
In the Standard Model, the higgs boson is introduced to explain the electroweak
symmetry breaking and the origin of fermion masses. 
The direct search at the 
CERN $e^{+}e^{-}$ collider (LEP2) presents a lower limit on the higgs boson 
mass of $m_{h}$ $>$ 114.4 GeV/$c^{2}$ at the 95$\%$ confidence level (C.L.).
Indirect measurements, electroweak global fits, give an upper limit of 144 
GeV/$c^{2}$ at the 95$\%$ C.L. which increases to 182 GeV/$c^2$ when 
the direct search is included~\cite{LEP}.

Our physics objective is to search for the low-mass fermiophobic
higgs and the high-mass Standard Model higgs boson using like-sign dilepton
events produced by 
\begin{eqnarray*}
qq^{\prime}\rightarrow W^{\pm}h\rightarrow W^{\pm}W^{\ast}W^{\ast}\rightarrow
\ell^{\pm}\ell^{\pm}+X.
\end{eqnarray*}
The relevant higgs mass regions are above 110 GeV/$c^{2}$ for the fermiophobic 
higgs where the branching fraction of $h\rightarrow W^{\ast}W^{\ast}$ 
supersedes that of $h\rightarrow \gamma\gamma$, and above 160 GeV/$c^{2}$ for 
the Standard Model higgs where the branching fraction of 
$h\rightarrow b\bar{b}$ is overtaken by this channel. 

The fermiophobic higgs boson has no coupling to the fermions.
Existence of the fermiophobic higgs could be an indication that the origin of
particle masses would be different for the bosons and the fermions.
Such a particle can also arise as a CP-even scalar $h^{0}$ in the two higgs 
doublet model (2HDM) type I. 
The model has seven degrees of freedom: the five particle masses 
($h^{0}$,$ H^{0}$, $A^{0}$, $H^{\pm}$) and two angles ($\alpha$, $\beta$). 
In the type I, the lightest CP-even scalar $h^{0}$ couples to a fermion 
proportionally to $\cos\alpha$, 
and the $h^{0}$ becomes a fermiophobic higgs when $\alpha=\pi/2$.

%------------------------------------------------------------------------------
% Data Samples and Event Selection
%------------------------------------------------------------------------------
\section{\label{sec:Data}Data Sample \& Event Selection}
This analysis is based on the data with an integrated luminosity of 
1.9 fb$^{-1}$ collected with the CDF-II detector between March 2002 
and May 2007.
Detailed descriptions of the CDF-II detector can be found in~\cite{CDF}.
The data are collected with inclusive lepton triggers that requires
an electron with transverse energy ($E_{T}$) $>$ 18 GeV or 
a muon with transverse momentum ($p_T$) $>$ 18 GeV/$c$.
Starting from the inclusive lepton datasets, we apply a number of event
selection criteria to obtain a baseline dilepton sample.
The events are required to have primary vertices
within the region to ensure well-defined measurement of collisions by the 
detector and to pass a cosmic-ray veto.
We select events 
at least one electron with $E_{T}>$ 20 GeV and $p_{T}>$ 10 GeV/$c$,
or muon with $p_{T}>$ 20 GeV/$c$, which is considered to be responsible for 
firing the triggers we have chosen, 
and at least one other electron with $E_{T}>$ 6 GeV and $p_{T}>$ 6 GeV/$c$,
or muon with $p_{T}>$ 6 GeV/$c$, as the baseline dilepton selection.
The leptons must be found in the central detector ($|\eta|<$ 1.1) and 
within fiducial regions of the sub-detectors.
They are also required to be
isolated in terms of the calorimeter cone-isolation, with a cone size of 
$R=$ 0.4, to be less than 2 GeV, where $R = \sqrt{(\Delta\eta)^{2}+
(\Delta\phi)^{2}}$ is the radius in the $\eta$-$\phi$ space,
$\eta=-\ln(\tan\theta/2)$ is the pseudorapidity, the $\theta$ is
the polar angle with respect to the proton beam direction, and 
$\phi$ is the azimuthal angle.
The leptons must have a well-measured track, found both in the outer
tracking chamber and the inner silicon detector, with the $z$ coordinate and
the impact parameter at the closest approach point to the beamline being
consistent with coming from the primary vertex.
We then apply a series of lepton identification cuts which impose various 
internal consistencies of information obtained from 
sub-detectors and require detector responses consistent with electrons or 
muons. If the electron is consistent with being due to a photon conversion as 
indicated by the presence of an additional nearby track, the electron is 
vetoed. 

For the exactly two-lepton events passing our selection above,
we explicitly require a cut to ensure that the two leptons are coming from
the same vertex, 
and also apply a dilepton mass cut ($M_{\ell\ell}>$ 12 GeV/$c^{2}$) 
and a $Z$-event veto.
We finally require the like-sign charge combination to complete the baseline
event selection.
The final cut to obtain the result is discussed in \S\ref{sec:Fin}.
%------------------------------------------------------------------------------
% Backgrounds
%------------------------------------------------------------------------------
\section{\label{sec:Bkg} Backgrounds }
Although the like-sign requirement is quite effective to suppress QCD and 
known electroweak processes, 
we expect that fake-lepton backgrounds and residual 
photon-conversions still remain at a considerable level in the events of our
signature.
They are 
estimated by data-driven methods, while other backgrounds containing prompt
real-leptons (physics backgrounds) are estimated by Monte Carlo (MC) data. 

\subsection{\label{sec:Phybkg} Physics Backgrounds }
The physics backgrounds can be classified into reducible and irreducible 
backgrounds. 
The reducible backgrounds are Drell-Yan, $WW$, $t\bar{t}$, 
and W + (heavy-flavor hadrons), 
while irreducible backgrounds are $WZ$ and $WW$. 
The reducible backgrounds are reduced by the 
isolation cut and like-sign requirement.
Contributions of the irreducible backgrounds are small due to their small
production cross sections, which are further suppressed by the $Z$-event veto.
Since fake-leptons and residual photon-conversions are estimated separately
using real data, we reject them found in the MC by looking at 
the generator-level information to avoid double counting.

\subsection{\label{sec:ResCo} Residual Photon-Conversion Background }
The residual photon-conversion backgrounds arise from an electron originating 
from the photon conversion with an unobserved partner track due to its low 
momentum. 
We estimate the backgrounds by multiplying lepton + conversion events by 
residual photon-conversion rate ($R_{{\rm res}}$). We define the rate by
\begin{eqnarray*}
R_{{\rm res}} = \frac{1-\varepsilon_{{\rm con}}}{\varepsilon_{{\rm con}}},
\end{eqnarray*} 
where $\varepsilon_{{\rm con}}$ is the conversion detection efficiency.
The efficiency is measured by comparing the conversions found in the real data
with conversion MC samples that are tuned to match with the sub-sample of
real conversions in the high-\pt\ region of partner-tracks 
where the efficiency is well known.
The residual conversion rate is parametrized by the parent photon \pt\ and
is shown in FIG.~\ref{fig:RCFR} left.
 
\subsection{\label{sec:Fakes} Fake Lepton Background }
The fake electron backgrounds are interactive $\pi^{\pm}$, overlap of $\pi^{0}$
and a track, and residual photon-conversions. The fake muon backgrounds are
punch-through hadrons and decay-in-flight muons from $\pi^{\pm}$ and $K^{\pm}$.
We also regard leptons from semileptonic decays of heavy-flavor
hadron as one of fake lepton backgrounds.
These background objects are common in generic QCD events.
We estimate the backgrounds by multiplying lepton + isolated track events by 
the fake lepton rates derived from inclusive jet samples. 
The fake lepton rate ($R_{\rm fake}$) is defined as a rate of leptons 
in the jet samples relative to isolated tracks with certain energy depositions
especially in the hadron calorimeters.
The \pt\ cut is 6 \pgev\ in accordance with the lower \pt\ cut of our baseline
event selection.
We reject $W$ and $Z$ events to find leptons in the jet samples to avoid
prompt real-leptons from electroweak processes.
The fake lepton rate is shown in FIG.~\ref{fig:RCFR} right.
The fake electron rates go through corrections to subtract residual conversions
in the jet samples
because we estimate the amount of residual photon-conversion events
separately as mentioned in the previous section.
    
%------------------------------------------------------------------------------
% Final Cut 
%------------------------------------------------------------------------------
\section{\label{sec:Fin} Final Cut }
Comparisons of 
kinematics distributions and the number of events between the data and the 
background prediction for the like-sign dilepton events after the baseline
selection are shown in
FIG.~\ref{fig:LSkine} and TABLE~\ref{tab:CDFvsBG},
where  
we categorize electron-muon pairs into $e\mu$ and $\mu e$ depending on 
which lepton has the leading \pt\ and fires the corresponding lepton trigger.

We further examine the baseline like-sign dilepton events 
on the 2 dimensional plane of the 2nd lepton $p_{T}$ ($p_{T2}$) versus 
the dilepton system $p_{T}$ ($p_{T12}$).
By scanning this 2D plane with moving $p_{T2}$ and $p_{T12}$ cut values, 
we see no significant discrepancies between the data and the background
expectations.
We then determine the optimized cut values for signal events to get 
minimized expected upper limit on the production cross section times
branching fraction (FIG.~\ref{fig:2D}).
The final cut is set to
$p_{T2}\geq$ 20 GeV/$c$ and $p_{T12}\geq$ 15 GeV/$c$.
With this cut, we define 4 regions:
\begin{itemize}
\item region1 : $p_{T2}\geq20$ GeV/$c$ and $p_{T12}<15$ GeV/$c$
\item region2 : $p_{T2}<20$ GeV/$c$ and $p_{T12}<15$ GeV/$c$
\item region3 : $p_{T2}<20$ GeV/$c$ and $p_{T12}\geq15$ GeV/$c$
\item region4 : $p_{T2}\geq20$ GeV/$c$ and $p_{T12}\geq15$ GeV/$c$
\end{itemize} 
TABLE~\ref{tab:CDFvsBG_2D} shows event counting for each region.
We see reasonable agreement between the expected number of backgrounds
and observed events.     
  
%------------------------------------------------------------------------------
% Acceptance and Systematic Uncertainties 
%------------------------------------------------------------------------------
\section{\label{sec:sys} Efficiency of Signal and Systematic Uncertainties } 
TABLE~\ref{tab:SigAccSys} (left) shows the signal efficiencies, where
the denominator is the number of $W^{\pm}h \rightarrow 
W^{\pm}W^{\ast}W^{\ast}\rightarrow \mbox{at least 2-lepton}$ events.
The branching fraction corresponding to events with at least 2 leptons from
triple $W$s is about 0.12.
The efficiencies are multiplied with MC scale factors which are obtained
by using Drell-Yan events found in the data and MC.
The uncertainties in TABLE~\ref{tab:SigAccSys} (right) 
include MC statistics and uncertainties of MC scale factors.
We estimate systematic uncertainties from several other sources for
the signal efficiencies as summarized in 
TABLE~\ref{tab:SigAccSys} (right).

Systematic uncertainties for the background estimation are already included
in TABLES~\ref{tab:CDFvsBG} and \ref{tab:CDFvsBG_2D}, 
which include uncertainties of residual photon-conversion estimation, 
fake-lepton estimation, and MC scale factors for the MC-based background
estimations.
        
%------------------------------------------------------------------------------
% Results 
%------------------------------------------------------------------------------
\section{\label{sec:lim} Results }
We find the expected number of backgrounds is 3.23 $\pm$ 0.69 events, while we 
observed 3 events in the data. 
From these results, 
we set the 95$\%$ C.L. upper limits on the cross section times branching 
fraction $\sigma(p\bar{p}\rightarrow Wh)\times Br(h\rightarrow W^{+}W^{-}$)
with the Bayesian approach as
2.2 pb for the higgs of the mass 110 GeV/$c^{2}$ and 1.4 pb for 160 GeV/$c^2$.
TABLE~\ref{tab:limit} shows the limit for each higgs mass point from 110 
GeV/$c^{2}$ to 200 GeV/$c^{2}$.
We also obtain the ratios of the limit to the Standard Model
cross sections and to the fermiophobic higgs cross sections (assuming the 
Standard Model production cross sections) with taking into account of
uncertainties of these theoretical predictions in the limit calculations, 
which are shown in TABLE~\ref{tab:limR}.
 
%------------------------------------------------------------------------------
% Conclusions
%------------------------------------------------------------------------------
\section{\label{sec:con} Conclusions }
We searched for the neutral higgs production associated with the $W$ boson
using high-$p_{T}$ like-sign dilepton events with the data corresponding to
an integrated luminosity of 1.9 ${\rm fb}^{-1}$.
We did not see any significant disagreements between background expectations 
and the data.
Given this, we tuned the final cut on the 2nd lepton $p_{T}$ and the $p_{T}$ of
dilepton system in order to obtain the limit-best result.
We found that the 2nd lepton $p_{T}$ cut of 20 GeV/$c$ and the dilepton-system
$p_{T}$ cut of 15 GeV/$c$ was a reasonable choice.
With this final cut, we obtained the expected number of fermiophobic higgs 
events for the mass 110 GeV/$c^{2}$ to be 0.46 assuming the Standard 
Model production cross section and 0.19 for the 160 GeV/$c^{2}$ higgs.
The expected number of backgrounds in the final sample was
$3.23 \pm 0.69$, while the actual number of observed events was 3.
We obtained the upper limits on the production cross-section times
the branching fraction for the higgs with masses in the region 
from 110 GeV/$c^{2}$ to 200 GeV/$c^{2}$, 
2.2 pb for the 110 GeV/$c^{2}$ higgs and 1.4 pb for 160 GeV/$c^{2}$. 

%------------------------------------------------------------------------------
% Acknowledgments
%------------------------------------------------------------------------------
\begin{acknowledgments}
We thank the Fermilab staff and the technical staffs of the
participating institutions for their vital contributions. This
work was supported by the U.S. Department of Energy and National
Science Foundation; the Italian Istituto Nazionale di Fisica
Nucleare; the Ministry of Education, Culture, Sports, Science and
Technology of Japan; the Natural Sciences and Engineering Research
Council of Canada; the National Science Council of the Republic of
China; the Swiss National Science Foundation; the A.P. Sloan
Foundation; the Bundesministerium fuer Bildung und Forschung,
Germany; the Korean Science and Engineering Foundation and the
Korean Research Foundation; the Particle Physics and Astronomy
Research Council and the Royal Society, UK; the Russian Foundation
for Basic Research; the Comision Interministerial de Ciencia y
Tecnologia, Spain; and in part by the European Community's Human
Potential Programme under contract HPRN-CT-20002, Probe for New
Physics.
\end{acknowledgments}

%------------------------------------------------------------------------------
% Tables 
%------------------------------------------------------------------------------
\clearpage
\input ./tables.fil

%------------------------------------------------------------------------------
% Figures
%------------------------------------------------------------------------------
\clearpage
\input ./figures.fil

\clearpage
%------------------------------------------------------------------------------
% Thebibliography
%------------------------------------------------------------------------------

\end{document}

%% file: tables.fil
% Background table ------------------------------------------------------------
\begin{table}[h]
\footnotesize
\centering
\caption{Background expectation and observed number of events for the 
like-sign dilepton events for each dilepton channel before the final cut.
}
\vspace{5mm}
\label{tab:CDFvsBG}
\begin{tabular}{lrrrrr}
\hline\hline
& $ee$ & $e\mu$ & $\mu e$ & $\mu\mu$ & Total \\
\hline
Fakes              & 46.1 &  29.8  & 30.4 & 19.8 & 126.1 \\
Photon-conversions &  6.2 & $-0.4$ &  6.4 & ---  &  12.2 \\
\hline
Total                       
&$52.3\pm 6.9$ &$29.4\pm 4.0$ &$36.8\pm 5.1$ &$19.7\pm 3.1$ &$138.3\pm 9.2$ \\
(Stat.)                     
&$\pm 2.2$     &$\pm 1.2$     &$\pm 1.7$     &$\pm 0.8$     &$\pm 3.1$ \\
(Syst.)                     
&$\pm 6.2$     &$\pm 3.8$     &$\pm 4.8$     &$\pm 3.0$     &$\pm 8.7$ \\
\\
$W \to e\nu/\mu\nu$        & 0.00 & 0.00 & 0.00 & 0.00 & 0.00 \\
$Z/\gamma^* \to ee/\mu\mu$ & 0.00 & 0.00 & 0.09 & 0.09 & 0.18 \\
$Z/\gamma^* \to \tau\tau$  & 0.00 & 0.00 & 0.00 & 0.00 & 0.00 \\
$t\overline{t}$            & 0.00 & 0.01 & 0.00 & 0.00 & 0.01 \\
Diboson                    & 0.46 & 0.65 & 0.79 & 0.82 & 2.58 \\
\hline
Total MC                    
&$0.46\pm 0.03$ &$0.66\pm 0.06$ &$0.88\pm 0.11$ &$0.91\pm 0.11$
&$2.86\pm 0.37$ \\
(Stat.)                      
&$\pm 0.01$     &$\pm 0.02$     &$\pm 0.09$     &$\pm 0.09$    
&$\pm 0.13$ \\
(Syst.)                      
&$\pm 0.03$     &$\pm 0.03$     &$\pm 0.05$     &$\pm 0.04$    
&$\pm 0.07$ \\
(Luminosity)                
&$\pm 0.03$     &$\pm 0.04$     &$\pm 0.05$     &$\pm 0.05$    
&$\pm 0.17$ \\ 

\hline
Fermiophobic higgs
&$0.227\pm 0.013$ &$0.224\pm 0.013$ &$0.271\pm 0.016$ &$0.214\pm 0.012$ 
&$0.938\pm 0.050$ \\
(110 GeV/$c^{2}$) & \mc{4}{c}{} \\
(Stat.)
&$\pm 0.004$ &$\pm 0.004$ &$\pm 0.004$ &$\pm 0.004$ &$\pm 0.009$ \\
(Syst.) 
&$\pm 0.003$ &$\pm 0.002$ &$\pm 0.002$ &$\pm 0.004$ &$\pm 0.014$ \\
(Luminosity)
&$\pm 0.013$ &$\pm 0.013$ &$\pm 0.016$ &$\pm 0.012$ &$\pm 0.050$ \\
\hline
SM higgs
&$0.063\pm 0.003$ &$0.061\pm 0.003$ &$0.084\pm 0.005$ &$0.064\pm 0.003$ 
&$0.272\pm 0.016$ \\
(160 GeV/$c^{2}$) & \mc{4}{c}{} \\
(Stat.)
&$\pm 0.001$ &$\pm 0.001$ &$\pm 0.001$ &$\pm 0.001$ &$\pm 0.002$ \\
(Syst.) 
&$\pm 0.001$ &$\pm 0.001$ &$\pm 0.001$ &$\pm 0.001$ &$\pm 0.003$ \\
(Luminosity)
&$\pm 0.003$ &$\pm 0.003$ &$\pm 0.005$ &$\pm 0.003$ &$\pm 0.016$ \\ 

\hline
Total expected              
&$52.8\pm 6.6$ &$30.6\pm 4.0$ &$37.6\pm 5.1$ &$20.7\pm 3.1$ &$141.7\pm 9.2$ \\
Data                        
&\mc{1}{c}{51} &\mc{1}{c}{26} &\mc{1}{c}{41} &\mc{1}{c}{16} &\mc{1}{c}{134} \\
\hline\hline
\end{tabular}
\end{table}

% Background table for 2D plane -----------------------------------------------
\begin{table}[h]
\footnotesize
\centering
\caption{Background expectation and observed number of events for the 
like-sign dilepton events for each region as defined in the texts. 
The region 4 correspods to the signal region after the final cut.
}
\vspace{5mm}
\label{tab:CDFvsBG_2D}
\begin{tabular}{lrrrrr}
\hline\hline
& \mc{4}{c}{Regions} \\
& 1 & 2 & 3 & 4 & Total \\
\hline
Fakes              &   0.47  &  27.8  & 96.5 &   1.14   & 125.9 \\
Photon-conversions &   0.52  &   0.5  & 10.5 &   0.70   &  12.2 \\
\hline
Total                       
&$0.99\pm 0.35$ &$28.3\pm 2.1$ &$107.0\pm 5.1$ &$1.84\pm 0.68$
&$138.1 \pm 6.1$ \\
(Stat.)                     
&$\pm 0.23$     &$\pm 1.8$     &$\pm 3.6$      &$\pm 0.30$     
&$\pm 4.1$ \\
(Syst.)                     
&$\pm 0.27$     &$\pm 1.1$     &$\pm 3.6$      &$\pm 0.61$     
&$\pm 4.5$ \\
\\
$W \to e\nu/\mu\nu$       & 0.00 & 0.00 & 0.00 & 0.00 & 0.00 \\
$Z/\gamma^* \to ee/\mu\mu$& 0.00 & 0.12 & 0.12 & 0.00 & 0.23 \\
$Z/\gamma^* \to \tau\tau$ & 0.00 & 0.00 & 0.00 & 0.00 & 0.00 \\
$t\overline{t}$           & 0.00 & 0.00 & 0.00 & 0.01 & 0.01 \\
Diboson                   & 0.11 & 0.06 & 1.05 & 1.38 & 2.60 \\
\hline
Total MC                    
&$0.11\pm 0.01$ &$0.17\pm 0.12$ &$1.16\pm 0.14$ &$1.40\pm 0.09$
&$2.84\pm 0.24$ \\
(Stat.)                      
&$\pm 0.01$     &$\pm 0.12$     &$\pm 0.12$     &$\pm 0.02$    
&$\pm 0.17$ \\
(Syst.)                      
&$\pm 0.00$     &$\pm 0.00$     &$\pm 0.01$     &$\pm 0.01$
&$\pm 0.02$ \\
(Luminosity)                
&$\pm 0.01$     &$\pm 0.01$     &$\pm 0.07$     &$\pm 0.08$
&$\pm 0.17$ \\ 

\hline
Fermiophobic higgs
&$0.036\pm 0.002$ &$0.014\pm 0.003$ &$0.418\pm 0.026$ &$0.464\pm 0.029$ 
&$0.932\pm 0.058$ \\
(110 GeV/$c^{2}$) & \mc{4}{c}{} \\
(Stat.)
&$\pm 0.001$ &$\pm 0.001$ &$\pm 0.006$ &$\pm 0.006$ &$\pm 0.009$ \\
(Syst.) 
&$\pm 0.000$ &$\pm 0.002$ &$\pm 0.006$ &$\pm 0.006$ &$\pm 0.014$ \\
(Luminosity)
&$\pm 0.002$ &$\pm 0.001$ &$\pm 0.025$ &$\pm 0.028$ &$\pm 0.056$ \\
\hline
SM higgs
&$0.011\pm 0.001$ &$0.002\pm 0.000$ &$0.075\pm 0.004$ &$0.183\pm 0.011$ 
&$0.272\pm 0.016$ \\
(160 GeV/$c^{2}$) & \mc{4}{c}{} \\
(Stat.)
&$\pm 0.000$ &$\pm 0.000$ &$\pm 0.001$ &$\pm 0.002$ &$\pm 0.002$ \\
(Syst.) 
&$\pm 0.000$ &$\pm 0.000$ &$\pm 0.001$ &$\pm 0.002$ &$\pm 0.003$ \\
(Luminosity)
&$\pm 0.001$ &$\pm 0.000$ &$\pm 0.004$ &$\pm 0.011$ &$\pm 0.016$ \\

\hline
Total expected              
&$1.10\pm 0.35$ &$28.5\pm 2.1$ &$108.1\pm 5.1$ &$3.23\pm 0.69$
&$141.0\pm 6.1$ \\
Data                        
&\mc{1}{c}{2} &\mc{1}{c}{23} &\mc{1}{c}{106} &\mc{1}{c}{3} &\mc{1}{c}{134} \\
\hline\hline
\end{tabular}
\end{table}

% Signal Acceptance and Systematic --------------------------------------------
\begin{table}[h]
\footnotesize
\caption{ Efficiencies for higgs events shown step by step for each higgs mass 
  (left) and  Systematic uncertainties of the signal efficiency (right).}
\label{tab:SigAccSys}
\begin{tabular}{cc}
\begin{minipage}{0.5\hsize}
\centering
\vspace{5mm}
\begin{tabular}{cccccc}
\hline\hline
Mass & Lepton ID & Dilepton cuts & Like-sign & $Z$ veto & Final cut \\
(\mgev) & (\%) & (\%) & (\%) & (\%) & (\%) \\
\hline
110 & 6.98 $\pm$ 0.10 & 6.77 $\pm$ 0.09 & 2.29 $\pm$ 0.03 & 2.11 $\pm$ 0.03 & 
1.05 $\pm$ 0.01 \\
120 & 7.41 $\pm$ 0.10 & 7.23 $\pm$ 0.10 & 2.43 $\pm$ 0.03 & 2.23 $\pm$ 0.03 & 
1.16 $\pm$ 0.02 \\
130 & 7.81 $\pm$ 0.11 & 7.66 $\pm$ 0.10 & 2.54 $\pm$ 0.04 & 2.33 $\pm$ 0.03 & 
1.33 $\pm$ 0.02 \\
140 & 8.26 $\pm$ 0.11 & 8.11 $\pm$ 0.11 & 2.68 $\pm$ 0.04 & 2.45 $\pm$ 0.03 & 
1.48 $\pm$ 0.02 \\
150 & 8.48 $\pm$ 0.11 & 8.35 $\pm$ 0.11 & 2.71 $\pm$ 0.04 & 2.45 $\pm$ 0.03 & 
1.57 $\pm$ 0.02 \\
160 & 8.41 $\pm$ 0.11 & 8.29 $\pm$ 0.11 & 2.63 $\pm$ 0.04 & 2.36 $\pm$ 0.03 & 
1.59 $\pm$ 0.02 \\
170 & 8.58 $\pm$ 0.11 & 8.46 $\pm$ 0.11 & 2.70 $\pm$ 0.04 & 2.43 $\pm$ 0.03 & 
1.67 $\pm$ 0.02 \\
180 & 9.08 $\pm$ 0.11 & 8.98 $\pm$ 0.11 & 2.85 $\pm$ 0.04 & 2.53 $\pm$ 0.03 &
1.79 $\pm$ 0.02 \\
190 & 9.55 $\pm$ 0.12 & 9.46 $\pm$ 0.12 & 3.12 $\pm$ 0.04 & 2.76 $\pm$ 0.04 &
1.98 $\pm$ 0.03 \\
200 & 9.76 $\pm$ 0.12 & 9.68 $\pm$ 0.12 & 3.25 $\pm$ 0.04 & 2.87 $\pm$ 0.04 & 
2.10 $\pm$ 0.03 \\
\hline\hline
\end{tabular}
\end{minipage}
\begin{minipage}{0.5\hsize}
\centering
\vspace{5mm}
\begin{tabular}{lcccc|c}
\hline\hline
Source & \mc{5}{c}{Uncertainty} \\
& $ee$ & $e\mu$ & $\mu e$ & $\mu\mu$ & $\ell\ell$ \\
& (\%) & (\%)   & (\%)    & (\%)     & (\%)       \\
\hline
Monte-Carlo statistics       &  2.6 &  2.4 &  2.1 & 2.2 & 1.6 \\
Monte-Carlo scale factors    &  1.1 &  1.0 &  1.3 & 1.2 & 1.2 \\
Parton distribution function &  1.9 &  1.8 &  2.1 & 2.0 & 1.9 \\
Initial state radiation      &  5.1 &  5.2 &  7.4 & 2.5 & 4.0 \\
Final state radiation        &  5.2 &  6.8 &  7.5 & 3.7 & 4.1 \\
$Z$ Cross Section            &  3.4 &  3.4 &  3.4 & 3.4 & 3.4 \\
Luminosity                   &  6.0 &  6.0 &  6.0 & 6.0 & 6.0 \\ \hline
Total                        & 10.7 & 11.7 & 13.1 & 9.0 & 9.6 \\
\hline\hline
\end{tabular}
\end{minipage}
\end{tabular}
\end{table}

% limit -----------------------------------------------------------------------
\begin{table}[h]
\centering
\caption{ Cross section upper limit at the 95$\%$ confidence level. }
\vspace{5mm}
\label{tab:limit}
\begin{tabular}{cccc}
\hline\hline
Mass (\mgev) & Expected Median (pb) & Expected Mean (pb) 
& Observed (pb)\\ \hline
110 & $2.2^{+0.4}_{-0.7}$ & 2.4 $\pm$ 0.8 & 2.2 \\
120 & $2.0^{+0.4}_{-0.7}$ & 2.2 $\pm$ 0.8 & 2.0 \\
130 & $1.8^{+0.3}_{-0.6}$ & 1.9 $\pm$ 0.7 & 1.8 \\
140 & $1.6^{+0.3}_{-0.5}$ & 1.7 $\pm$ 0.6 & 1.6 \\
150 & $1.5^{+0.3}_{-0.5}$ & 1.6 $\pm$ 0.6 & 1.5 \\
160 & $1.5^{+0.3}_{-0.5}$ & 1.6 $\pm$ 0.5 & 1.5 \\
170 & $1.4^{+0.3}_{-0.5}$ & 1.5 $\pm$ 0.5 & 1.4 \\
180 & $1.3^{+0.3}_{-0.4}$ & 1.4 $\pm$ 0.5 & 1.3 \\
190 & $1.2^{+0.2}_{-0.4}$ & 1.3 $\pm$ 0.4 & 1.2 \\
200 & $1.1^{+0.2}_{-0.4}$ & 1.2 $\pm$ 0.4 & 1.1 \\
\hline\hline
\end{tabular}
\end{table}

% limit ratio -----------------------------------------------------------------
\begin{table}[h]
\caption{ Ratio of the cross section upper limit to the Standard Model 
  prediction (left) and to the fermiophobic higgs. }
\vspace{5mm}
\label{tab:limR}
\begin{tabular}{cc}
\begin{minipage}{0.5\hsize}
\centering
\begin{tabular}{cccc}
\hline\hline
Mass (\mgev) & Expected Median & Expected Mean  
& Observed \\ \hline
110 & $281.8^{+57.1}_{-92.9}$ & 305.7 $\pm$ 111.2 & 281.8 \\
120 & $115.7^{+23.4}_{-38.1}$ & 125.5 $\pm$  45.6 & 115.7 \\
130 & $ 60.2^{+12.2}_{-19.8}$ &  65.3 $\pm$  23.7 &  60.2 \\
140 & $ 40.8^{+ 8.2}_{-13.4}$ &  44.3 $\pm$  16.1 &  40.8 \\
150 & $ 34.9^{+ 7.0}_{-11.5}$ &  37.8 $\pm$  13.7 &  34.9 \\
160 & $ 33.1^{+ 6.3}_{-10.9}$ &  35.9 $\pm$  13.0 &  33.1 \\
170 & $ 38.3^{+ 7.7}_{-12.6}$ &  41.5 $\pm$  15.1 &  38.3 \\
180 & $ 47.2^{+ 9.5}_{-15.5}$ &  51.2 $\pm$  18.6 &  47.2 \\
190 & $ 54.1^{+10.9}_{-17.8}$ &  58.7 $\pm$  21.3 &  54.1 \\
200 & $ 64.2^{+13.0}_{-21.1}$ &  69.6 $\pm$  25.3 &  64.2 \\
\hline\hline
\end{tabular}
\end{minipage}
\begin{minipage}{0.5\hsize}
\centering
\begin{tabular}{cccc}
\hline\hline
Mass (\mgev) & Expected Median & Expected Mean  
& Observed \\ \hline
110 & $12.9^{+ 2.6}_{- 4.2}$ & 14.0 $\pm$  5.1 & 12.9 \\
120 & $15.7^{+ 3.2}_{- 5.2}$ & 17.1 $\pm$  6.2 & 15.7 \\
130 & $18.4^{+ 4.3}_{- 6.0}$ & 20.0 $\pm$  7.2 & 18.4 \\
140 & $21.5^{+ 5.3}_{- 7.1}$ & 23.4 $\pm$  8.5 & 21.5 \\
150 & $26.3^{+ 6.3}_{- 8.6}$ & 28.6 $\pm$ 10.4 & 26.3 \\
160 & $31.4^{+ 6.5}_{-10.3}$ & 34.0 $\pm$ 12.4 & 31.4 \\
170 & $37.9^{+ 7.7}_{-12.4}$ & 41.1 $\pm$ 14.9 & 37.9 \\
180 & $47.2^{+ 9.5}_{-15.5}$ & 51.2 $\pm$ 18.6 & 47.2 \\
190 & $54.1^{+10.9}_{-17.8}$ & 58.7 $\pm$ 21.3 & 54.1 \\
200 & $64.2^{+13.0}_{-21.1}$ & 69.6 $\pm$ 25.3 & 64.2 \\
\hline\hline
\end{tabular}
\end{minipage}
\end{tabular}
\end{table}

%% file: figures.fil
%  Residual conversion rate and Fake lepton rate ==============================
\begin{figure}[h]
\centering 
\hspace{-5mm}
\includegraphics[width=0.33\textwidth]{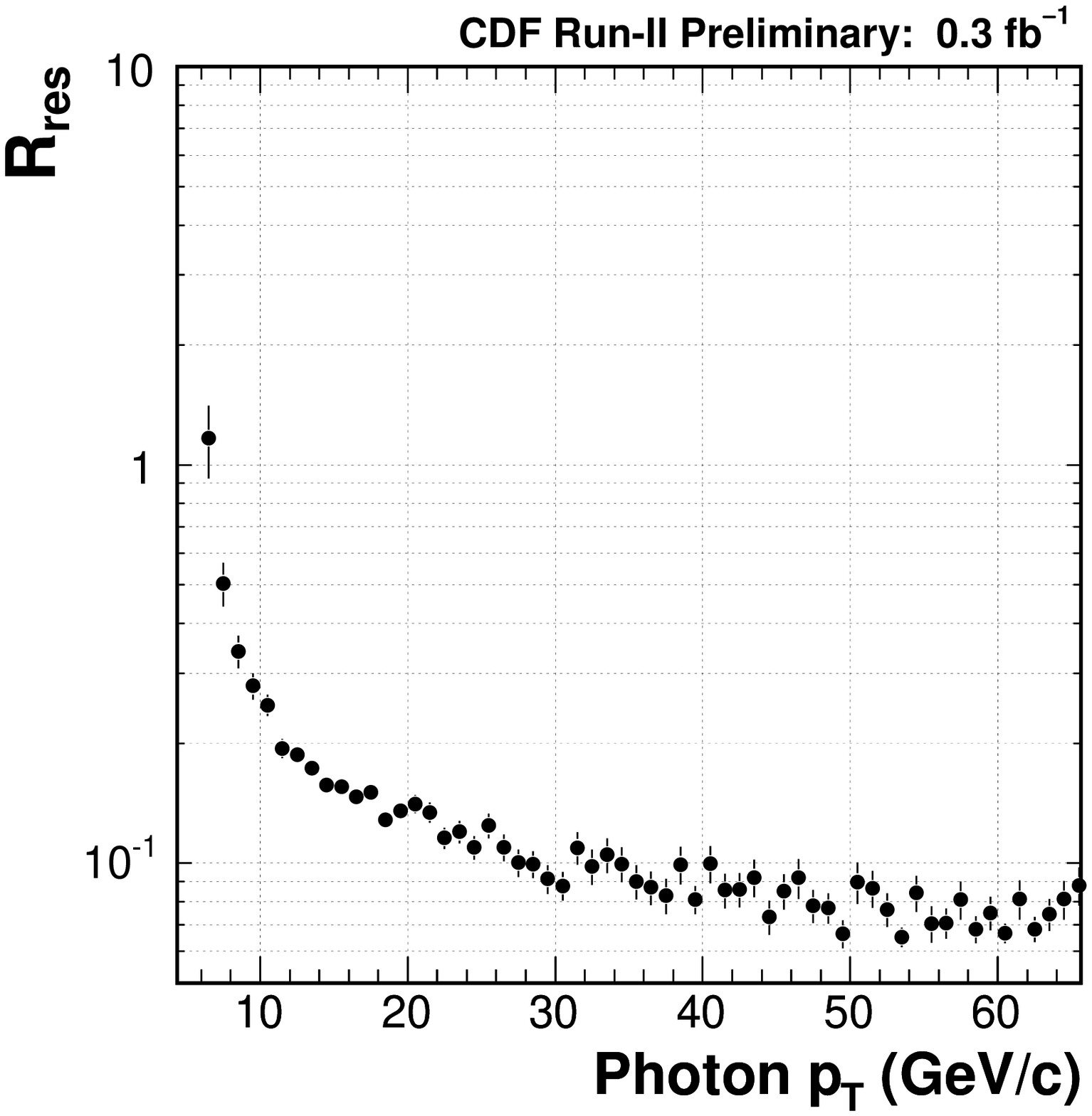}
\hspace{-5mm}
\includegraphics[width=0.33\textwidth]{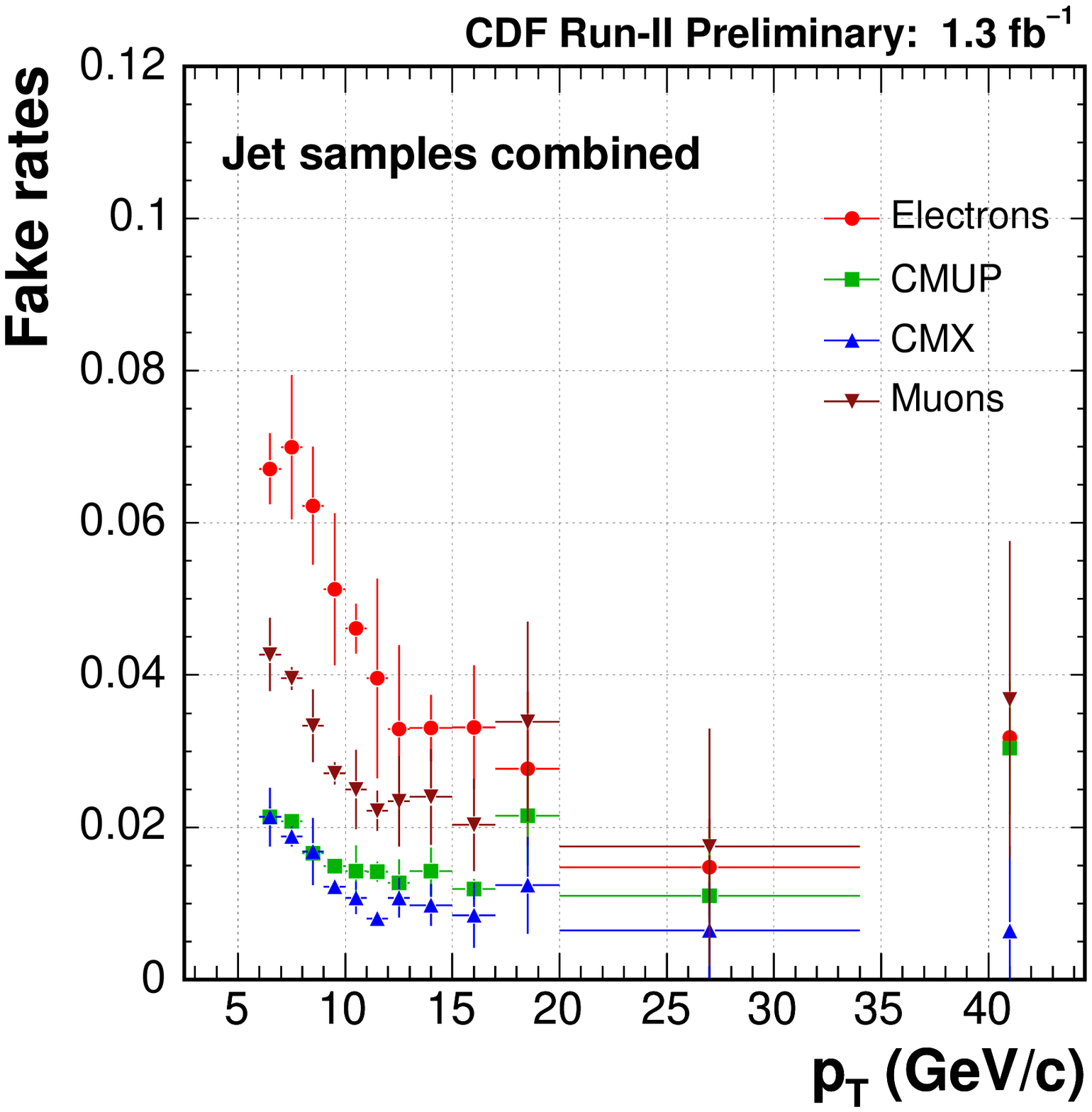}
\caption{ Residual photon-conversion rates as a function of photon $p_{T}$ 
  (left) and   Fake-lepton rates as a function of isolated track $p_{T}$ 
  for each lepton object(right).}
\label{fig:RCFR}
\end{figure}

% LS kinematics (pT12) ========================================================
\begin{figure}[h]
\centering 
\hspace{-5mm}
\includegraphics[width=0.33\textwidth]{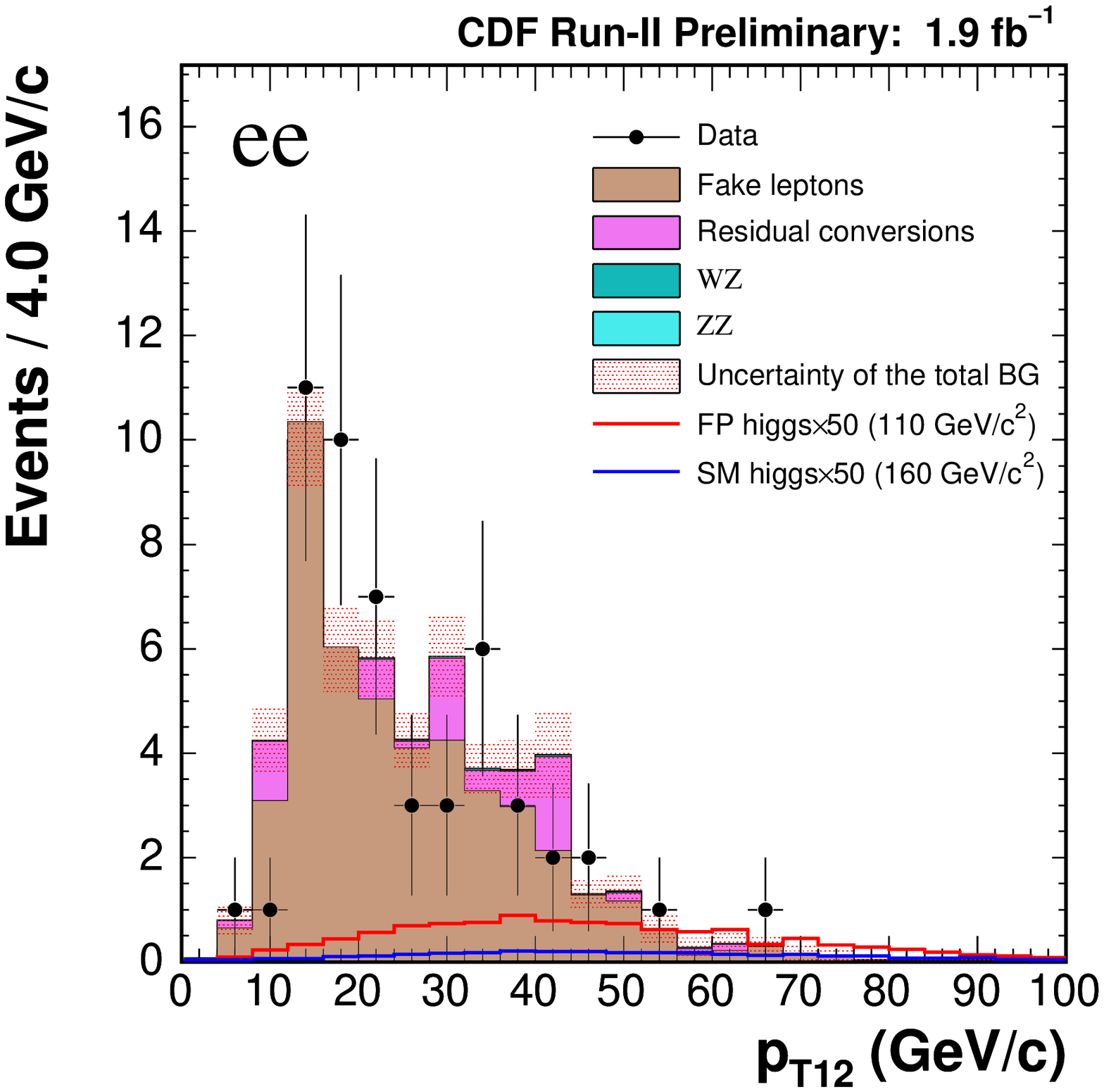}
\hspace{-5mm}
\includegraphics[width=0.33\textwidth]{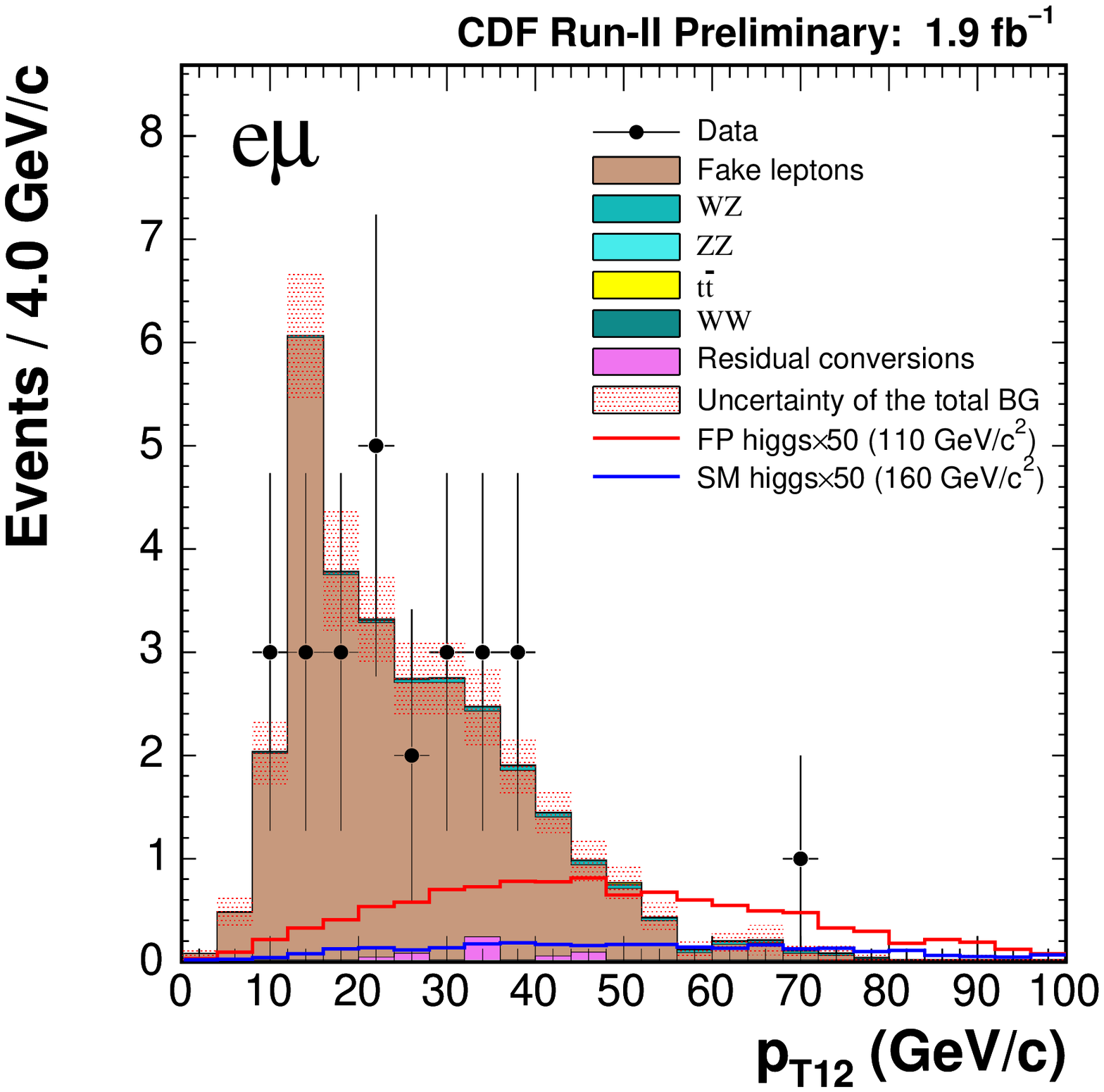}
\\
\hspace{-5mm}
\includegraphics[width=0.33\textwidth]{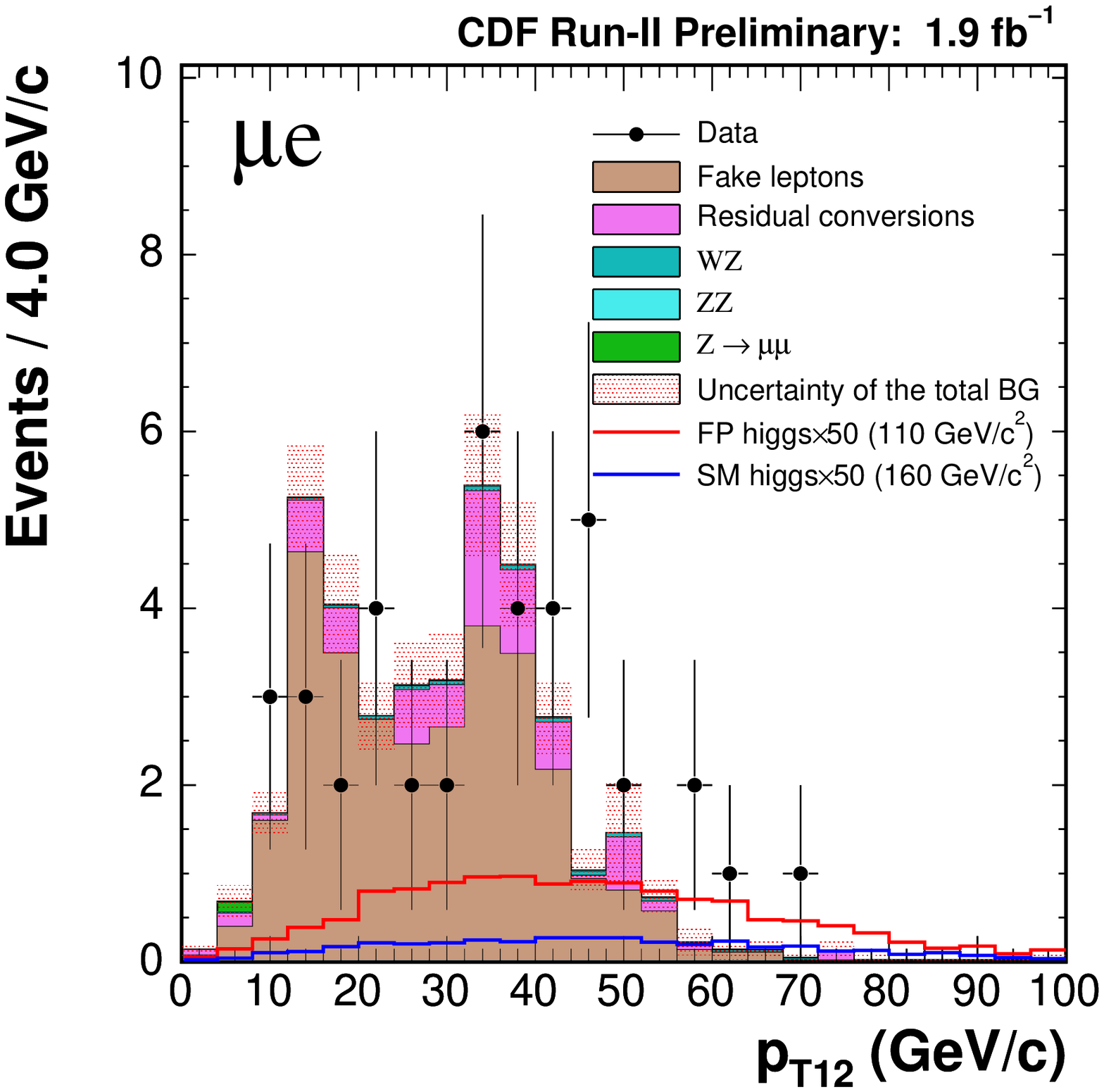}
\hspace{-5mm}
\includegraphics[width=0.33\textwidth]{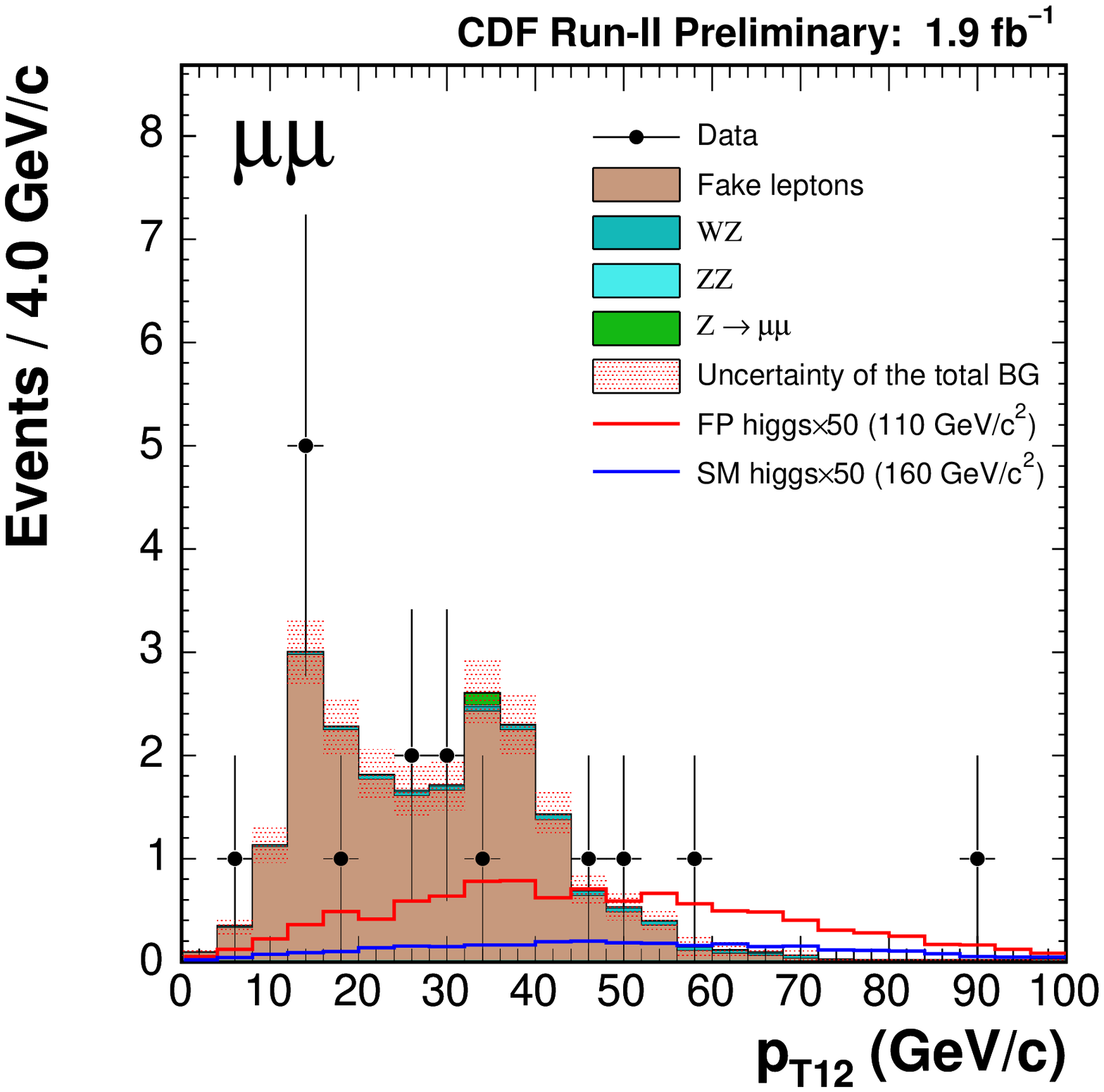}
\caption{ Comparison of kinematic distributions (dilepton system $p_{T}$) 
  between the data and the background prediction for the like-sign dilepton 
  events. }
\label{fig:LSkine}
\end{figure}

% Expected limit on the 2D plane ==============================================
\begin{figure}[h]
\centering 
\hspace{-5mm}
\includegraphics[width=0.5\textwidth]{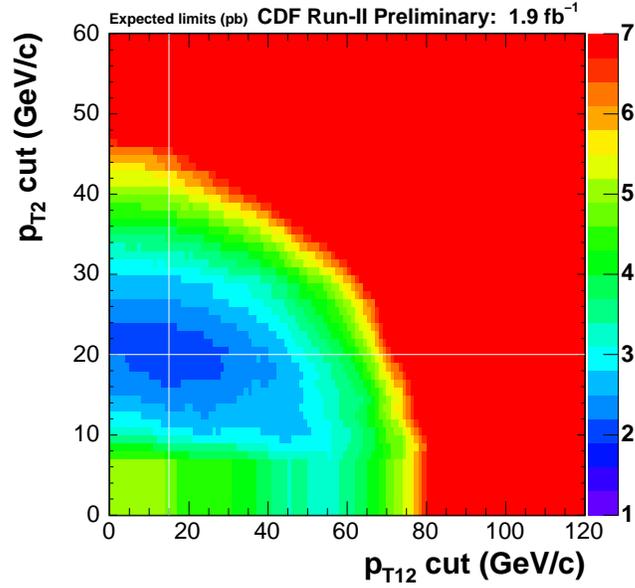}
\caption{ Expected upper limit on the 
$\sigma(Wh)\times Br(h\rightarrow W^{+}W^{-})$ at the 95$\%$ confidence level 
on the 2 dimensional plane. }
\label{fig:2D}
\end{figure}

%  95% CL upper limit =========================================================
\begin{figure}[h]
\centering 
\hspace{-5mm}
\includegraphics[width=0.33\textwidth]{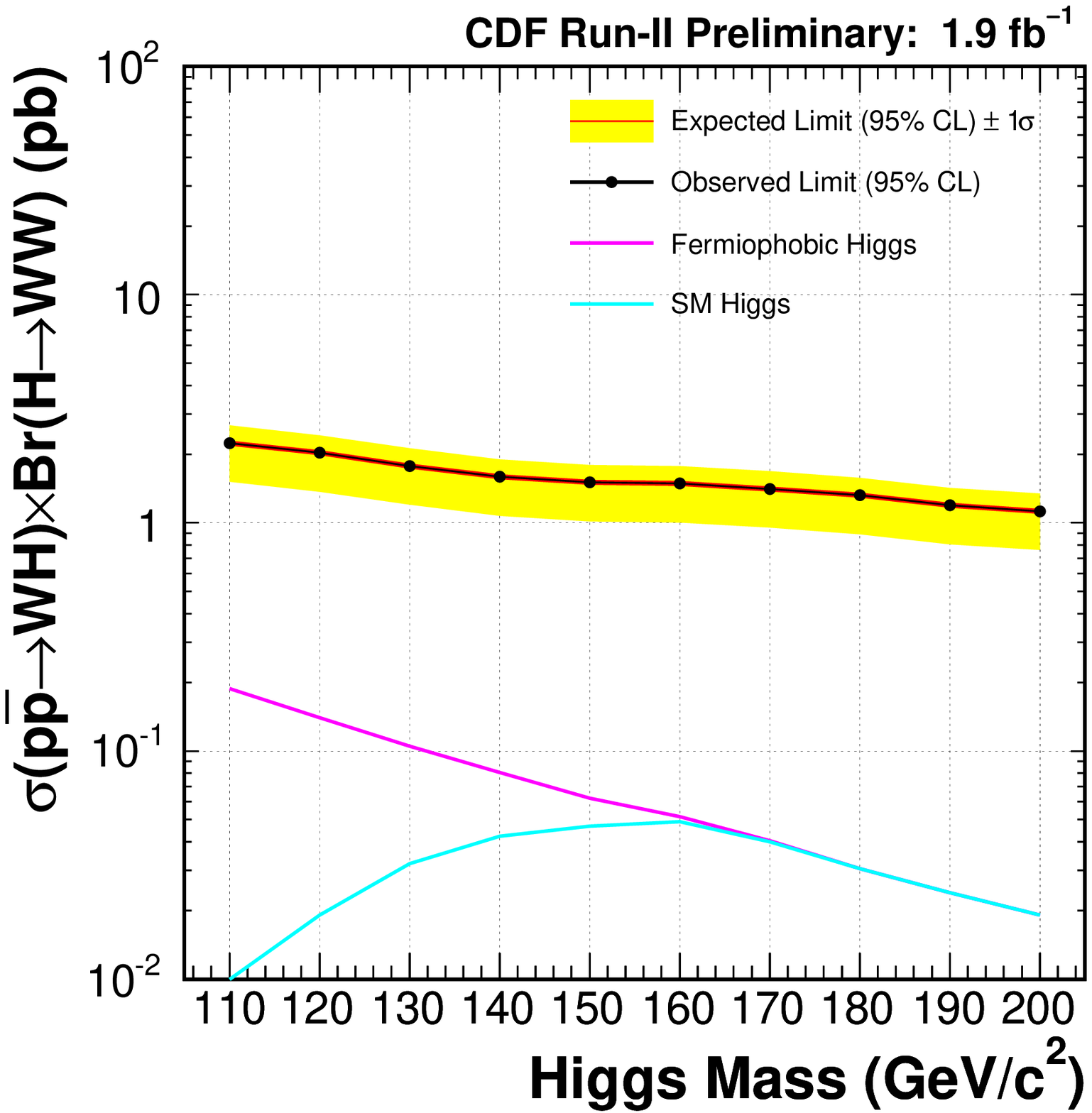}
\hspace{-5mm}
\includegraphics[width=0.33\textwidth]{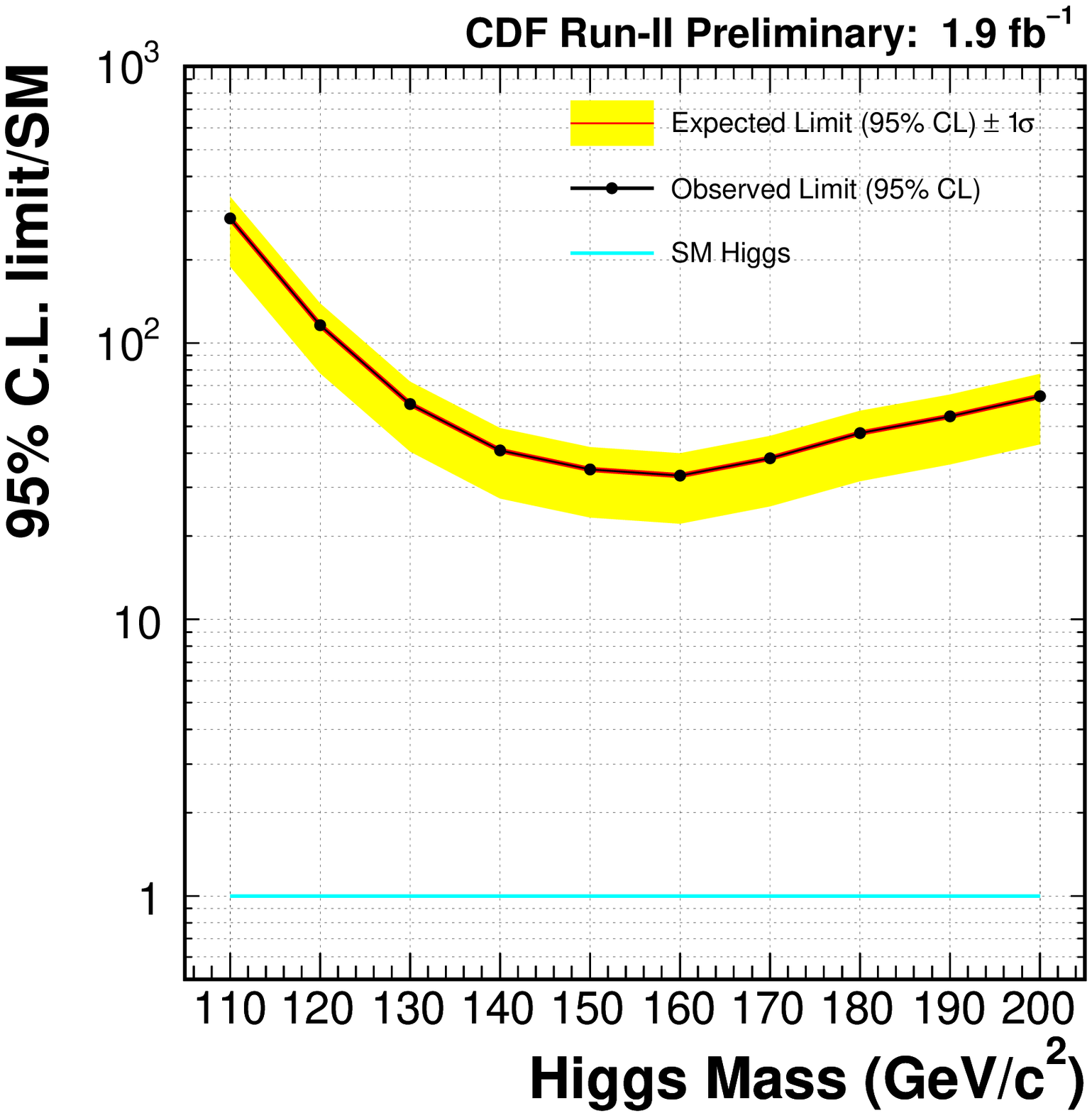}
\hspace{-5mm}
\includegraphics[width=0.33\textwidth]{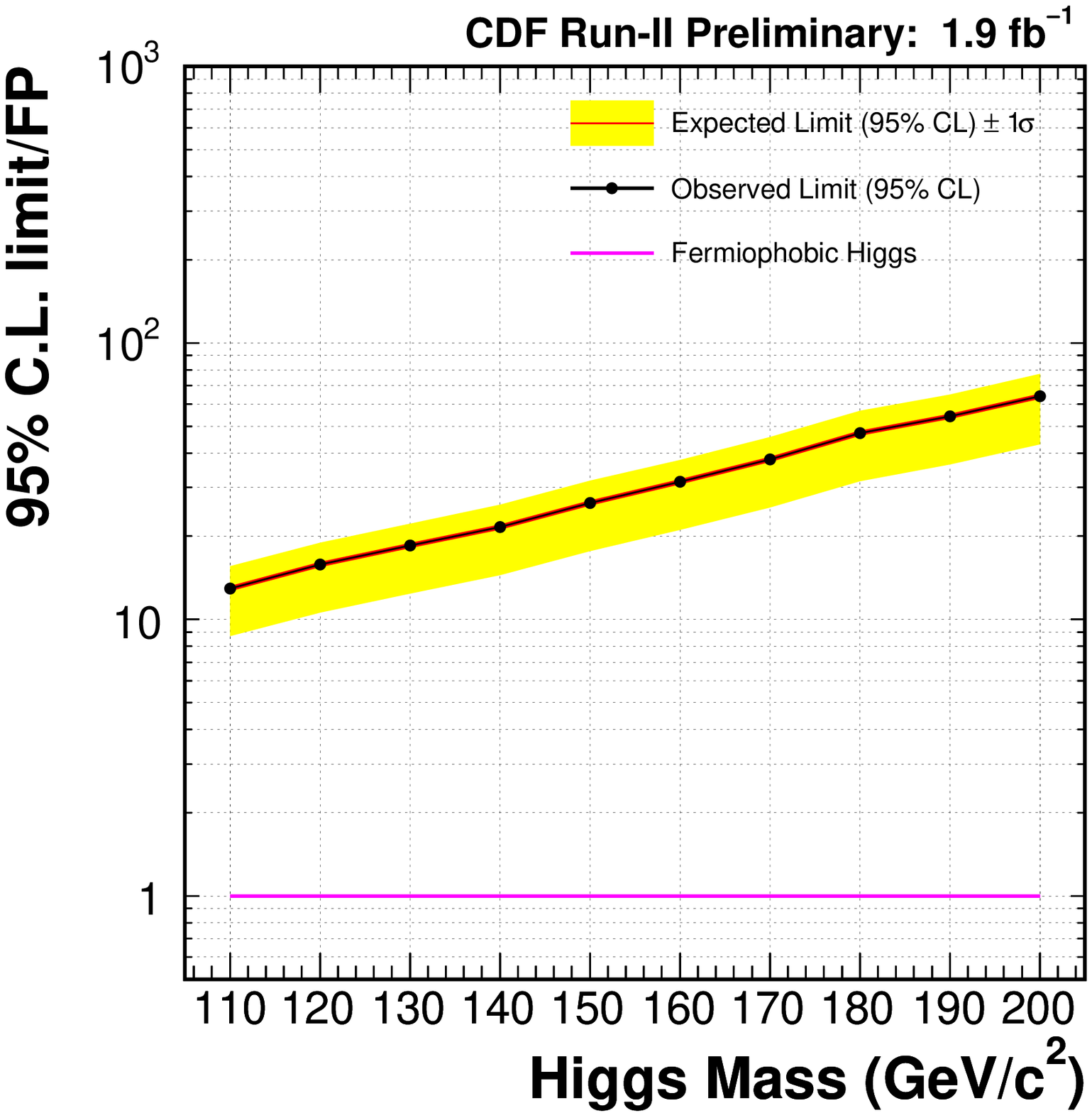}
\caption{ Cross section upper limit at the 95$\%$ confidence level (left) and 
  Ratio of cross section upper limit to the prediction of the Standard Model 
  (center) and fermiophobic higgs (right). }
\label{fig:limit}
\end{figure}